\documentstyle [11pt,epsf] {article}
\title{Explosion Implosion Duality and the Laboratory Simulation of
Astrophysical Systems}

\author{L O'C Drury \\
Dublin Institute for Advanced Studies\\
5 Merrion Square\\
Dublin 2\\
Ireland
\and 
J T Mendon\c{c}a\\
Grupo de Lasers e Plasmas \\
Instituto Superior T\'echnico \\
Lisboa 1096 CODEX\\
Portugal}

\date{24 March 2000}
\def\be{\begin{equation}}
\def\ee{\end{equation}}
\def\Ed{{\cal E}}
\def\ddt#1{{\partial#1\over\partial t}}

\def\DDt#1{{D#1\over D\, t}}
\def\tSW{t_{\rm SW}}

\begin{document}

\maketitle

\begin{abstract}
The Euler equations of ideal gas dynamics posess a remarkable
nonlinear involutional symmetry which allows one to factor out an
arbitrary uniform expansion or contraction of the system. The nature
of this symmetry (called by cosmologists the transformation to
supercomoving variables) is discussed and its origin clarified.  It is
pointed out that this symmetry allows one to map an explosion problem
to a dual implosion problem and vice versa. The application to
laboratory simulations of supernova remnants is considered; in
principle this duality allows the complete three-dimensional evolution
of highly structured explosion ejecta to be modelled using a static
target in an implosion facility.

\end{abstract}

\section{Introduction}
There is much interest at present in the possible use of the new
generation of high-power laser facilities (in particular the National
Ignition Facility at Livermore and the Laser MegaJoule in Bordeaux)
to simulate astrophysical phenomena such as supernovae.  At first
sight this programme appears to suffer from one obvious drawback. The
phenomena one wishes to simulate generally involve {\em explosions}
while the laser facilities are designed to produce {\em
implosions}. Remarkably, as we will show, this is not a problem. Under
certain, not too restrictive, conditions there exists an exact
mathematical duality which allows one to transform an explosion
problem to an implosion problem and vice versa. Thus it is possible,
in a precise sense, to use {\em implosion} experiments to simulate
{\em exploding} systems.

\section{The duality transformation}

The Euler equations of perfect gas dynamics can be conveniently written in
the form
\begin{eqnarray}
\DDt\rho &=& -\rho \nabla\cdot{\bf U}\\
\DDt {{\bf U}} &=& - {\nabla p \over \rho}\\
\DDt\Ed &=& - (\Ed + p) \nabla\cdot {\bf U}
\end{eqnarray}
where $D/Dt$ denotes the Lagrangian, material or convective derivative
defined by
\be
\DDt{} = \ddt{} + {\bf U} \cdot \nabla
\ee
and $\rho$ is the mass density, $\Ed$ the thermal energy density, $p$ the
pressure and $\bf U$ the velocity. In addition to these differential
equations one algebraic relation is needed, an equation of state relating
the pressure to the mass and energy densities the simplest being a
polytropic equation of state,
\be
p = (\gamma-1) \Ed
\ee
with $\gamma$ a constant.
These equations are mathematically equivalent to the mass, momentum and
energy conservation equations in smooth regions of the flow. Only at
shocks is it necessary to revert to the fundamental conservation forms
to recover the correct shock jump conditions. 

Now consider the following transformation of the dependent and
independent variables,
\begin{eqnarray}
{\bf x}^*   &=& a(t)^{-1} {\bf x}, \\
t^*         &=& \int a(t)^{-2} dt, \\
\rho^*      &=& a(t)^3 \rho, \\
p^*         &=& a(t)^5 p, \\
{\bf U}^*   &=& a(t) {\bf U} - \dot a(t) {\bf x}, \\
\Ed^*       &=& a(t)^5 \Ed
\end{eqnarray}
where for the moment $a(t)$ is an arbitrary function of time. 
Apart from the, at first sight rather strange, time-dependent scaling factors
this is essentially a transformation to a coordinate system which is expanding or contracting with a scale factor $a(t)$. If we  define $a^* = a^{-1}$ and note that
\be
{d a^*\over d t^*} = a^2 {d\over dt}\left(1\over a\right) = - {d a\over dt}
\ee
it is easy to see that it is an involutionary transformation 
with inverse obtained by simply interchanging the starred and unstarred 
quantities
\begin{eqnarray}
{\bf x}   &=& a^*(t^*)^{-1} {\bf x}^*, \\
t         &=& \int a^*(t^*)^{-2} dt^*, \\
\rho      &=& a^*(t^*)^3 \rho^*, \\
p         &=& a^*(t^*)^5 p^*, \\
{\bf U}   &=& a^*(t^*) {\bf U}^* - \dot a^*(t^*) {\bf x}^*, \\
\Ed       &=& a^*(t^*)^5 \Ed^*.
\end{eqnarray}

Let us now consider how the dynamical equations transform under this
change of variables. It is easy to see that
\begin{eqnarray}
\DDt{} &\to& a^{*2} {D\over D t^*} \\
\nabla &\to& a^* \nabla^* 
\end{eqnarray}
and thus, after some elementary algebra,
\begin{eqnarray}
{D \rho^*\over D t^*} &=& - \rho^* \nabla^* \cdot {\bf U}^*, \\
{D {\bf U}^*\over D t^*} &=& - {\nabla^* p^*\over \rho^*} 
+ {{\ddot a}^*\over a^*} {\bf x}^*, \\
{D \Ed^*\over D t^*} &=& - (\Ed^* + p^*) \nabla^*\cdot {\bf U}^*
+ { {\dot a}^* \over a^* } ( 3 p^* - 2 \Ed^*).
\end{eqnarray}
Remarkably, we see that if the scale factor $a$ is such that \be
{\ddot a}^* = {d^2 a^*\over d t^{*2}} = - a^2 \ddot a = 0 \ee and the
gas is a polytrope of exponent $5/3$ with $p = 2\Ed/3$ then the Euler
equations are {\em invariant} under this transformation.  Note that,
because the Euler equations in conservation form are algebraically
equivalent to the simplified forms, the conservation forms are also
invariant and thus the whole structure of ideal gas dynamics, 
including the Rankine-Hugoniot shock relations, is preserved.

The condition that the acceleration of the scale factor be zero,
$\ddot a = 0$, requires that $a(t)$ be a linear function of $t$ and,
without loss of generality, we can take $a = t/t_0$ where $t_0$ is a
constant characteristic expansion time. The time transformation is
then \be t^* = \int {dt\over a(t)^2} = t_0^{2}\int{dt\over t^2} =
\hbox{const.} - {t_0^{2}\over t} \ee and it is convenient to set
the constant to zero and choose \be t^* = -{t_0^2\over  t}, \qquad t
= -{t_0^2\over t^*}.  \ee The initial singularity of the expansion
in physical space occurs at $t=0$ and is mapped to $t^* = -\infty$,
the long term behaviour as $t\to +\infty$ is mapped to $t^* = 0$. It
is important to note that in the dual representation the time variable
is bounded from {\em above}, $t^* < 0$, whereas in physical space it is
bounded from {\em below}, $t>0$.

The remarkable result is that for an ideal gas of point particles with
no internal structure (which is what the 5/3 polytrope is)
hydrodynamics in a uniformly expanding system is exactly equivalent to
hydrodynamics in a static system. This result, or special forms of it,
appears to have been discovered a number of times by Cosmologists
(where the idea of factoring out the general expansion of the universe
is very natural); a recent discussion is that of Martel and Shapiro
(1998) where they propose the felicitous name of ``supercomoving
variables'' to describe this transformation.  What does not seem to
have been generally noted is that this transformation can be used
outside the cosmological context (however Poyet and Spiegel, 1979, did
use a variant in an analysis of stellar pulsations).

\section {Interpretation}

The fact that the transformation is exact for the gas of ideal point
particles strongly hints that it is derived from a similar result for
the free particle motion. In fact there is such a duality, although it
is almost trivial. The freely moving point particle moves along a
straight line trajectory, \be {\bf x} = {\bf x}_0 + {\bf v}_0 t, \ee
with starting point ${\bf x}_0$ and velocity ${\bf v}_0$. If we write
this as \be {{\bf x}\over t} = {\bf v}_0 + {\bf x}_0 {1\over t} \ee we
see that there is a dual representation of the trajectory in which $t$
is replaced by $1/t$, lengths are scaled by a factor proportional to
time, initial points and final velocities are interchanged, but the
trajectory remains a straight line. If collisions are instantaneous,
localised and elastic they look the same in either system, and thus in
both systems one can write down a Boltzmann equation and then derive
the hydrodynamic equations as limits of moments of the Boltzmann
equation. This approach also shows that higher order effects, such as
viscosity and heat conduction, can formally be treated in the same
way; however the resulting transformed transport coefficients will in
general have unphysical time dependencies (for an application see
Drury and Stewart, 1976).

This analysis also shows that similar results will hold in different
numbers of spatial dimensions, but the equation of state will have to
correspond to the ideal gas in that number of dimensions. In $d$
spatial dimensions it is easy to verify that the "super-comoving"
transformation takes the form
\begin{eqnarray}
{\bf x}^*   &=& a(t)^{-1} {\bf x}, \\
t^*         &=& \int a(t)^{-2} dt, \\
\rho^*      &=& a(t)^d \rho, \\
p^*         &=& a(t)^{d+2} p, \\
{\bf U}^*   &=& a(t) {\bf U} - \dot a(t) {\bf x}, \\
\Ed^*       &=& a(t)^{d+2} \Ed
\end{eqnarray}
and that the Euler equations are invariant if the gas has a polytropic equation of state such that
\be
d p = 2 \Ed
\ee
corresponding to an adiabatic exponent
\be
\gamma = 1 + {2\over d}.
\ee

An interesting way of looking at this transformation (for which we are
indebted to our colleague Etienne Parizot) is that it provides an
analogue in spherical geometry to the freedom that Galilei transformations
allow in planar geometry. If we are looking at a planar shock, it is
often convenient to transform to a reference frame where the upstream
medium, or the downstream medium, or the shock itself, appears
stationary. In spherical systems one cannot apply Galilei boosts
because the origin is fixed, however this transformation, by allowing
one to take out an arbitrary uniform expansion, gives one much the
same freedom.

\section{Application to a Supernova Remnant}

Computational studies of the evolution of a Supernova Remnant commonly
start with initial conditions of dense pressure-free ejecta expanding
ballistically away from the site of the explosion, which it is
convenient to locate at the coordinate origin, and interacting with a
stationary, or slowly moving, ambient medium of much lower density and
negligible pressure.  To illustrate the application of the duality
transformation let us consider the simple, if somewhat artificial,
case of uniform density ejecta interacting with a uniform and
stationary ambient medium in perfect spherical symmetry. Then the
initial conditions correspond to
\begin{eqnarray}
\rho(r,t) &=& \rho_0 \left(t\over \tSW\right)^{-3}, \\
U(r,t) &=& {r\over t},\\
p(r,t) &=& 0
\end{eqnarray}
in the region $r<V_0 t$ occupied by the ejecta ($V_0$ is the maximum
expansion speed of the ejecta) and
\begin{eqnarray}
\rho(r,t) &=& \rho_0\\
U(r,t) &=& 0, \\
p(r,t) &=& 0
\end{eqnarray} in the external ($r \gg V_0 t$)
medium of constant density $\rho_0$.  The sweep-up time $\tSW$
corresponds to the point where the ejecta, if expanding unimpeded,
would have a density equal to the ambient medium.

This defines the physical problem of expanding ejecta interacting with
a stationary environment. Let us now consider the dual problem
obtained by applying the transformation with scale factor
\be
a (t) = {t\over \tSW}.
\ee
Then the dependent variables transform as
\begin{eqnarray}
r^* = \tSW {r\over t},\\
t^* = - {\tSW^2\over t}
\end{eqnarray}
so that the explosion, which occurs at $t=0$ in physical problem, occurs
at $t^*=-\infty$ in the dual problem. Conversely the asymptotic
evolution as $t\to\infty$ in the physical problem is mapped to the
behaviour at $t^* = 0$ in the dual problem.

The ejecta density in the dual problem is constant,
\be
\rho^*(r^*, t^*) = a^3 \rho(r,t) = \rho_0
\ee
and the velocity is zero, $U^* = 0$, in $r^* < V_0 \tSW$. 
However the ambient medium is now time-dependent with density, in the
region $r^* \gg V_0\tSW$
\be
\rho^*(r^*, t^*) = \left(t\over\tSW\right)^3 \rho_0 =
\left(-t^*\over\tSW\right)^{-3} \rho_0
\ee
and velocity
\be 
U^*(r^*, t^*) =  {r^*\over t^*}.
\ee

Thus in the dual problem we have {\em stationary} ejecta
interacting with an {\em imploding} ambient medium whereas in the
physical problem we have {\em exploding} ejecta interacting with
a {\em stationary} ambient medium. Instead of the initial explosion at
$t=0$ in the physical problem we have the final crunch at $t^*=0$ in
the dual problem. 

The evolution in physical space of the supernova remnant structure has
been often discussed and is well-known (eg Truelove and McKee, 1999;
Dwarkadas and Chevalier, 1998). At early times, $t\ll\tSW$, the bulk
of the ejecta expand ballistically except for a thin interaction
region on the outside consisting of a forward shock running into the
ambient medium, a zone of hot shocked ambient medium, a contact
discontinuity, a zone of shocked ejecta and a reverse shock
propagating slowly into the ejecta. At later times, when the mass of
swept up ambient material becomes comparable to the ejecta mass, the
reverse shock detaches itself from the contact discontinuity and
implodes on the centre and the outer forward shock approximates the
self-similar Sedov solution for a strong point explosion in a cold
gas.

In the dual system the interaction looks a little different, and in
some ways is simpler.  Initially we have the stationary sphere of high
density material (which for convenience we continue to call the
ejecta, although in the dual representation it has not been ejected
but is simply sitting there) surrounded by a very low density
converging flow. The inflowing gas has to decelerate at a shock which
stands about 10\% further out in radius than the edge of the
ejecta. Writing for convenience $\tau = - t^*/\tSW$ there is an exact
similarity solution in which $U^*\propto \tau^{-1}$, $\rho^*\propto
\tau^{-3}$ and $p^*\propto \tau^{-5}$ in the region external to the
sphere of ejecta. This steeply rising pressure ($\propto\tau^{-5}$)
drives the reverse shock into the ejecta and starts the implosion of
the ejecta.

\begin{figure}
\epsfxsize=0.9\hsize
\epsfbox{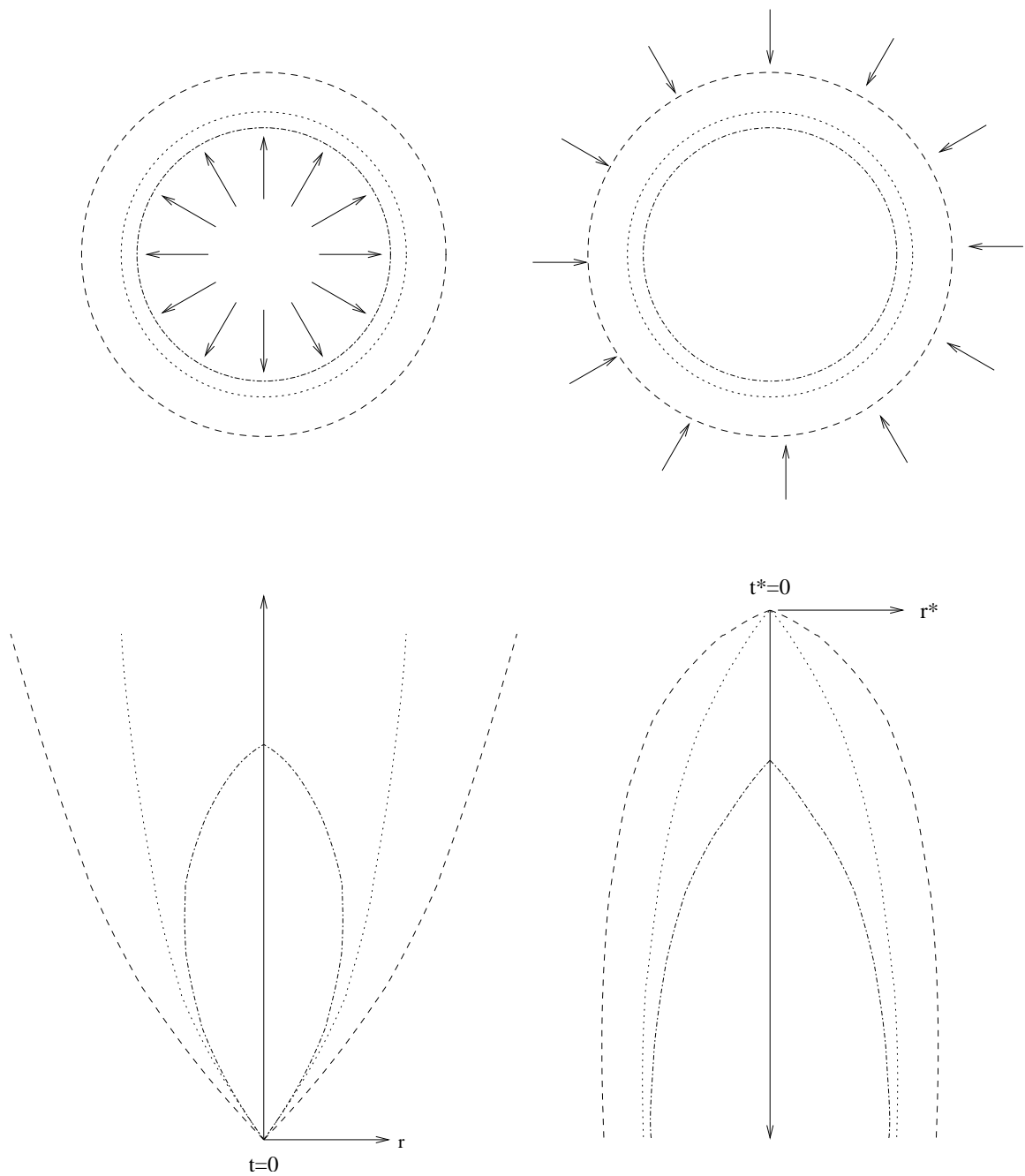}
\caption{A schematic representation of the SNR structure and evolution
as seen in physical space (left) and the dual space (right).  The
locations of the forward shock (dashed) , contact discontinuity (dotted) 
and reverse shock (dash-dotted) are indicated.}
\end{figure}

At later times, as the ejecta collapse, the shock in the imploding
ambient medium also moves inwards thereby reducing the rate of
increase of the pressure. Transforming the Sedov solution to the dual
system we see that the shock radius scales as \be r^* \propto
\tau^{3/5} \ee and the postshock pressure as \be p^*\propto
\tau^{-19/5}. \ee Figure~1 attempts to show schematically the relation
between the two representations. We note in passing that the dual 
representation is also useful for analytic and numerical studies;
this aspect will be explored in a companion paper (Dwarkadas and Drury,
in preparation).

\section{Prospects for laboratory simulations}

The perfectly symmetric explosion is neither realistic nor especially
interesting; it is the easiest case to analyse numerically and there
is no reason to suppose that a laboratory simulation would yield any
additional information. However reality is more complicated. It is
clear that the ejecta emerging from real supernova explosions are
highly nonuniform on a wide range of scales and that to calculate the
resulting remnant evolution in three dimensions is likely to remain a
computationally challenging problem for some considerable time (cf
Arnett, 1999).

The interesting implication of this work is that it should be possible
with the new generation of implosion facilities to simulate precisely
this problem, the interaction of highly structured ejecta with their
surroundings including all the effects of spherical geometry. One can
easily imagine constructing a solid target whose density distribution
models the density distribution of the expanding ejecta. If this
target is then used in an implosion experiment, and if the momentum
loading on the surface is tailored to rise in the same manner as the
pressure behind the forward shock in the dual system, a steep initial
rise as $\tau^{-5}$ decreasing to $\tau^{-3.8}$, the evolution of the
internal structures including all the turbulent mixing, instabilities
and shock formation, should be exactly replicated.

We emphasise finally that the transformation discussed in this paper
is additional to and complements the well-known linear scaling
relations as excellently discussed by Ryutov et al (1999) in the
astrophysical context, or Connor and Taylor (1977) in the plasma
physics context. Dimensional similarity and scaling arguments are
obviously central to any attempt at simulation on a laboratory scale
of astrophysical systems, however precisely because they are very
general and linear they cannot turn an explosion into an
implosion. The remarkable nonlinear symmetry discussed in this paper
is specific to the ideal gas equation of state, but subject to this
constraint gives a powerful new degree of freedom in simulation
studies by allowing an arbitrary uniform expansion or contraction to
be factored out thereby transforming an explosion problem to an
implosion one or vice versa.

\section {Acknowledgments}

This work was in part supported by the EU under the TMR programme,
contract FMRX-CT98-0168.  Some of it was carried out while LD was a
visitor at the Research Centre for Theoretical Astrophysics of the
University of Sydney.  

\section{References}

{
\parindent = 0 pt
\parskip = 5 pt plus 5 pt
\frenchspacing

Arnett, D. 1999, astro-ph/9909031

Connor J W and Taylor J B, 1977, Nuclear Fusion 17, 1047.

Drury, L O'C and Stewart, J M, 1976 MNRAS 177, 377.

Dwarkadas, V V and Chevalier, R A, 1998, ApJ 497, 807.

Martel, H and Shapiro, P R, 1998, MNRAS 297, 467.

Ryutov, D et al, 1999, ApJ 518, 821.

Truelove, J K and McKee, C F, 1999, ApJS 120, 299.

Poyet, J P and Spiegel, E A, 1979, AJ 84, 1918.

}
\end{document}